\documentclass[prb,aps,amssymb,twocolumn,superscriptaddress,notitlepage]{revtex4-2}
\usepackage{amsmath,esvect}
\usepackage{amssymb}
\usepackage{amsthm}
\usepackage{amsfonts}
\usepackage{listings}
\usepackage{physics}
\usepackage{enumerate}
\usepackage{latexsym}
\usepackage{psfrag}
\usepackage{bm}
\usepackage{graphicx}
\usepackage[caption=false]{subfig}
\usepackage{blkarray}
\usepackage{array}
\usepackage{color}
\usepackage{xcolor}
\usepackage[normalem]{ulem}
\usepackage{hyperref}
\hypersetup{
     colorlinks = true,
     linkcolor = magenta,
     citecolor = magenta
}

\def\beq{\begin{equation}}
\def\eeq{\end{equation}}
\def\bal{\begin{aligned}}
\def\eal{\end{aligned}}

\begin{document}

\title{Extrinsic Geometry and Gappable Edges in Rotationally Invariant Topological Phases}
\author{Sounak Sinha}
\affiliation{Department of Physics, University of Illinois Urbana-Champaign, Urbana IL 61801, USA}
\affiliation{Anthony J. 
Leggett Institute for Condensed Matter Theory, University of Illinois Urbana-Champaign, Urbana IL 61801, USA}

\author{Barry Bradlyn}
\email{bbradlyn@illinois.edu}
\affiliation{Department of Physics, University of Illinois Urbana-Champaign, Urbana IL 61801, USA}
\affiliation{Anthony J. 
Leggett Institute for Condensed Matter Theory, University of Illinois Urbana-Champaign, Urbana IL 61801, USA}

\date{\today}

\begin{abstract}
Recent work on Abelian topological phases with rotational symmetry has raised the question of whether rotational symmetry can protect gapless propagating edge modes. 
Here we address this issue by considering the coupling of topological phases to the extrinsic geometry of the background. 
First, we analyze an effective hydrodynamic theory for an Abelian topological phase with vanishing Hall conductance. 
After integrating out the bulk hydrodynamic degrees of freedom, we identify charge neutral, rotationally invariant mass terms by coupling the propagating boundary modes to the extrinsic geometry. 
This allows us to integrate out the edge modes and we find a gapped theory described by a local induced action that depends on the extrinsic geometry of the boundary, regardless of the shift. 
Finally, we apply these ideas to a microscopic theory and find the explicit bulk terms which respect gauge and rotational symmetry and open a gap in the edge spectrum.
\end{abstract}
\maketitle

\section{Introduction}

One of the main experimental diagnostics of correlated topological phases is their edge state dynamics~\cite{kane2005quantum,kane2005topological,bernevig2006quantum,konig2007quantum,hsieh2009observation,xu2012observation,xia2009observation,fu2007topological,fu2007topologicala,hsieh2012topological,teo2008surface}. 
Edge states arise due to the bulk-boundary correspondence, which can be interpreted from the point of view of anomaly inflow~\cite{arouca2022quantum,callan1985anomalies}. 
Topological phases enriched with or protected by a continuous symmetry group can have nonvanishing, nondissipative bulk responses to external fields that couple to the symmetry~\cite{bradlyn2015lowenergy,gromov2015thermal,wen1991topological,stone2012gravitational}. 
This leads to an accumulation of symmetry charge on boundaries of the system, which must be absorbed by some gapless excitations on the boundary. 
From the complementary point of view, the gapless boundary modes when viewed in isolation are described by an anomalous theory: the symmetry appears to be broken on the boundary due to quantum effects. 
Only when the boundary theory is coupled to the bulk is the symmetry restored and the total theory rendered non-anomalous. 
A similar story holds also for topological phases with discrete symmetries, though in that case the anomaly manifests itself in the failure of the partition function of the boundary theory to be gauge invariant, and the gauge symmetry is restored by a bulk topological action related to the cohomology of the symmetry group~\cite{witten2016three,dijkgraaf1990topological}. 

This picture works well when the symmetry group is an on-site symmetry. 
However, for crystalline symmetries, additional complications may arise. 
If the boundary does not respect the symmetries that protect a topological phase, gapless edge states need not appear~\cite{po2017symmetrybased,song2018quantitative,bradlyn2017topological,song2020realspace,slager2013space,kruthoff2017topological,cano2021band}. 
The recent discovery of ``higher-order'' topological insulators has revealed that the absence of gapless boundary modes does not always imply the absence of bulk topology~\cite{benalcazar2017quantized,benalcazar2017electric,schindler2018higher,khalaf2018symmetry,wieder2018axion}. 
These higher-order topological insulators are protected by bulk symmetries such as spatial inversion or discrete rotational symmetry that do not leave points on the boundary fixed, and for realistic systems do not even map the boundary into itself~\cite{schindler2018higherordera,lin2022spin,wieder2020strong,schindler2022topological}. 
Nevertheless, even though this allows the boundary modes to be gapped, the nontrivial bulk topology manifests in a boundary response that cannot be observed in an isolated system.

While such topological phases with gapped edge states are well-understood in noninteracting models, the situation for correlated phases is less clear. 
Recent advances such as the topological crystal method~\cite{song2019topological,huang2017building}, the crystalline equivalence principle~\cite{thorngren2018gauging}, and crystalline gauge theories~\cite{huang2022effective,manjunath2021crystalline,may2022crystalline} provide rigorous approaches to classifying the bulk topology of systems protected by or enriched with crystal symmetries. 
When it comes to edge modes, however, crystalline symmetries present a complication not present in most approaches to the classification of topological phases.  In particular, the embedding of the boundary of the system in space allows us to define an \emph{extrinsic} geometry of the boundary. 
Since the boundary is acted on by spatial symmetries, this extrinsic geometry couples directly to both the edge degrees of freedom and to gauge fields and defects associated to the crystal symmetry~\cite{rao2023effective}. 
This represents physics that goes beyond the crystalline equivalence principle, and so raises the question of what happens to edge states protected by crystalline symmetries. 

For the case of rotational symmetry in two dimensions, several recent works~\cite{gromov2016boundary,rao2023effective,manjunath2023rotational,liu2019shift} have shown that edge states of topological phases protected by discrete rotational symmetry can be gapped, leaving behind a boundary with a nontrivial response to curvature. 
In the simplest cases, this manifests as a fractional charge bound to corners of the system. 
For topological phases protected by continuous rotational symmetry, however, the fate of edge excitations is still unknown. 
It has recently been shown that in the absence of coupling to extrinsic geometry, the boundary theory of such states remains anomalous, and no mass term can exist that gaps the edge modes~\cite{manjunath2023rotational}. 

In this work, we revisit this question with a focus on the role of extrinsic geometry. 
We show that for abelian topological phases in two dimensions enriched with continuous rotational symmetry, the rotational symmetry alone cannot protect gapless edge excitations. 
To show this, we adapt the approach of Ref.~\cite{levin2009fractional} which showed how to construct mass terms that gap edge excitations given certain algebraic conditions on the modes. 
A naive application of these rules would imply that when the edge excitations are nontrivially charged under the rotational symmetry, no rotationally invariant mass term can be written. 
Here, however, we show explicitly how coupling between the edge modes and the extrinsic curvature allows for a Higgs-like mass to appear, where the normal and tangent vectors to the boundary play the role of a Higgs field. 
We show how to construct such mass terms, and show by explicit calculation that the low-energy field theory for the boundary of a 2D topological phase with rotational symmetry is given by the gapped ``Gromov-Jensen-Abanov'' action introduced in Ref.~\cite{gromov2016boundary}. 

The remainder of this work is organized as follows. 
First, in Sec.~\ref{sec:gjareview} we review intrinsic and extrinsic geometry of 2D systems, as well as the induced action for topological phases coupled to geometry. 
Next, in Sec.~\ref{sec:kmatrix} we derive the boundary action for an abelian topological phase charged under both electromagnetic and rotational symmetry, using an effective $K$-matrix Chern-Simons theory in the bulk. 
We pay particular attention to local counterterms needed to render the bulk and boundary action invariant under the symmetries, and show how the boundary degrees of freedom emerge and couple to the background geometry. 
In Sec.~\ref{sec:effection_action} we explicitly construct rotational and electromagnetic gauge-invariant mass terms for the boundary degrees of freedom. 
We show that by coupling the boundary modes to extrinsic geometry, there exist ``dressed'' neutral vertex operators in the boundary theory whenever a mutually commuting set of boundary modes exists, provided the Hall conductivity and chiral central charge of the topological phase vanish. 
Crucially, rotational symmetry places no constraint on the existence of these mass terms. 
We explicitly integrate out the boundary modes in the presence of the mass terms and show how the ``Gromov-Jensen-Abanov'' local counterterm emerges. 
Finally, in Sec.~\ref{sec:example} we demonstrate how these considerations apply to a concrete model of a (noninteracting) topological phase protected by rotational symmetry. 
Starting  from the microscopic model of a Dirac electron with nontrivial orbital spin first considered in Ref.~\cite{may2022crystalline}, 
we solve for the boundary modes on a general 2D manifold with boundary, with a focus on the coupling to extrinsic geometry. 
We show that there exists a rotationally invariant mass term written in terms of the bulk fermions that when projected into the Hilbert space of boundary modes opens a gap. 
Using bosonization, we derive that the charge response of the gapped boundary is given by the Gromov-Jensen-Abanov action. 
Finally, we conclude in Sec.~\ref{sec:outlook} by discussing the implications of our results for topological phases protected by discrete symmetries and for phases in three dimensions

\section{Review of The Gromov-Jensen-Abanov Action}\label{sec:gjareview}

Let us first review how 2D topological phases couple and respond to background geometry. 
We will assume that our $2+1$-dimensional spacetime manifold can be written as $\mathcal{M}=\mathbb{R}\times\Gamma$, where $\Gamma$ is a spatial manifold with boundary, and $\mathbb{R}$ is the time dimension. 
The boundary of our system lives on a $1+1$-dimensional spacetime manifold $\partial\mathcal{M}=\mathbb{R}\times \partial\Gamma$, which can be viewed as a domain wall between our topological phase and a trivial phase. 
Our primary objective in this work is to describe the (possibly gapped) degrees of freedom that are present on the boundary when the bulk phase has nontrivial topology and nontrivial coupling to the geometry of the manifold. 
We will describe the geometry of $\mathcal{M}$ using a first-order formalism. 
We introduce coordinates $X=(t,\mathbf{X})$ on $\mathcal{M}$, where $t$ is the time coordinate. 
We introduce a set of orthonormal frame fields (vielbeins)
\begin{equation}\label{eq:framedef}
    e^A = e^A_i \mathrm{d}X^i,
\end{equation}
where $i=1,2$ index coordinates on $\Gamma$, and $A=1,2$ is a flat internal index. 
The frame fields determine the metric tensor on $\Gamma$ via
\begin{equation}\label{eq:metricdef}
g_{ij} = \delta_{AB}e_i^Ae_j^B,  
\end{equation}
where here and throughout this work the Einstein summation convention is used. 
We also introduce the dual inverse vielbeins $E_A^i$ satisfying
\begin{align}
    E_A^ie_j^A &= \delta^i_j \\
    e_i^AE_B^i &=\delta^A_B.
\end{align}
Vielbeins at different points on the manifold are related via the (Levi-Civita) spin connection $\omega = \omega_t \mathrm{d}t + \omega_i\mathrm{d}X^i$. 
For the geometries under consideration, the spin connection can be expressed in terms of the vielbeins as
\begin{align}
\label{eq:spinconnection}
\omega_t &= \frac{1}{2}\epsilon^A_{\hphantom{A}B}E^i_A\partial_0 e_{i}^B, \nonumber  \\
\omega_k &= \frac{1}{2}\epsilon^A_{\hphantom{A}B}E^i_A\partial_k e_{i}^B + \frac{1}{2}\epsilon^{AB}E^i_A E^j_B\partial_i g_{jk}.
\end{align}
We see that the geometry defined by Eq.~\eqref{eq:metricdef} is invariant under an $SO(2)$ rotation by space-time dependent angle $f$,
\begin{align}\label{eq:introtsym}
    e^A&\rightarrow R^A_{\hphantom{A}B}(f)e^B \nonumber\\ 
    \omega &\rightarrow \omega+\mathrm{d}f,
\end{align}
which we refer to as internal rotational symmetry.

The frame fields and the spin connection define the intrinsic geometry of the manifold $\mathcal{M}$. 
However, on the boundary $\partial\mathcal{M}$ we have additional \emph{extrinsic} geometric structure. 
We can define a (spacelike) tangent vector $t^i$ to the boundary, as well as an (inward-pointing) normal one-form $n_i$ such that $n_i t^i = 0$. 
Taken together, $n_i$ and $t^i$ define the extrinsic geometry of the spatial boundary $\partial\Gamma$. 
We can define the extrinsic curvature
\begin{equation}
    K_{\bar{\mu}} = n_i D_{\bar{\mu}} t^i
\end{equation}
where $D_{\bar{\mu}}$ is the covariant derivative, and $\bar{\mu}=0,1$ is a boundary spacetime index. 
The Gauss-Bonnet theorem provides a relation between the extrinsic curvature and the spin connection. 
In particular, we can integrate over a spatial slice to find
\begin{equation}\label{eq:gaussbonnet}
    \int_\Gamma\mathrm{d}\omega + \int_{\partial\Gamma} K = 2\pi\chi(\Gamma)
\end{equation}
where $\chi(\Gamma)$ is the (generalized) Euler characteristic of $\Gamma$. 
For simple closed boundaries of the type we will consider here, $\chi(\Gamma)=1$. 
In a gauge where $\omega$ is defined over all of $\Gamma$, we can further use Stokes's theorem to deduce that
\begin{equation}\label{eq:alphadef}
    \omega|_{\partial\mathcal{M}}+K=\mathrm{d}\alpha,
\end{equation}
where $\mathrm{d}\alpha$ is a closed form with winding $2\pi\chi(\Gamma)$, and $\omega|_{\partial M}$ is the spin connection restricted to the boundary. 
Physically, $\alpha(X)$ measures the change in angle between the bulk frames parallel transported around the boundary, measured relative to the boundary normal and tangent vectors.

For gapped phases of matter in 2+1 dimensions coupled to electromagnetic fields and background geometry, we can integrate out the microscopic degrees of freedom to obtain an induced action that encodes the response of the system to the external fields (i.e. generates correlation functions). 
The energy gap ensures that the induced action will be a local functional of the electromagnetic gauge field $A$, the spin connection $\omega$, and also the frames $e^A$. 
The induced action must also respect both electromagnetic gauge symmetry $A\rightarrow A+\mathrm{d}g$ and the internal rotational symmetry Eq.~\eqref{eq:introtsym}. 
The lowest-order topological terms in the induced action satisfying these constraints in the bulk of $\mathcal{M}$ can be written as
\begin{equation}\label{eq:topactionbulk}
    S_{\mathrm{top}}=\int \frac{\sigma}{4\pi} A\wedge\mathrm{d}A + \frac{\bar{s}}{2\pi}A\wedge\mathrm{d}\omega + \frac{l_s}{4\pi}\omega \wedge\mathrm{d}\omega
\end{equation}
Here $\sigma$ is the Chern number and gives the Hall conductance of the phase (in units of the flux quantum), $\bar{s}$ is the ``orbital spin density'' and is proportional to the Hall viscosity, and $l_s$ is the ``orbital spin variance.'' We will restrict our attention to non-chiral topological phases so that the chiral central charge $c=0$, and so there is no separate gravitational Chern Simons term in Eq.~\eqref{eq:topactionbulk}. 
The action Eq.~\eqref{eq:topactionbulk} is gauge and internal rotationally invariant in the bulk, but not in the presence of a boundary. 
A nonzero $\sigma$ implies a gauge anomaly on the boundary, which necessitates the presence of gapless boundary modes. 
Naively, it also appears that when $\bar{s}$ is nonzero gauge symmetry is broken on the boundary, and when $l_s$ is nonzero internal rotational symmetry is broken on the boundary. 
This raises the question of whether anomaly inflow arguments require the existence of gapless modes even when $\sigma=0$.

In Ref.~\cite{gromov2016boundary}, it was shown however that there is no anomaly associated with nonzero $\bar{s}$ or nonzero $l_s$. 
In particular, they showed that there exist local counterterms on the boundary
\begin{align}\label{eq:gjaintro}
S_{\mathrm{GJA}}&=\int_{\partial\mathcal{M}} \frac{\bar{s}}{2\pi}A\wedge K + \frac{l_s}{4\pi} \omega\wedge K, \nonumber \\
&\equiv S_{\mathrm{GJA}_1}+S_{\mathrm{GJA}_2}
\end{align}
which couples the gauge fields to extrinsic geometry. 
The GJA action \eqref{eq:gjaintro} transforms under electromagnetic and internal rotational gauge symmetry to precisely cancel the variation of the bulk action Eq.~\eqref{eq:topactionbulk}. 
The combined action $S_\mathrm{top}+S_\mathrm{GJA}$ is local and has no anomalies beyond the gauge anomaly dictated by $\sigma$. 
This suggests that nonzero $\bar{s}$ and nonzero $l_s$ do not alone protect gapless edge modes.

Nevertheless, questions remain about how the GJA action arises, and how modes at the boundary of a topological phase can be explicitly coupled to and gapped by the extrinsic geometry when $\sigma=0$. 
For cases where the internal rotational symmetry is broken to a discrete subgroup, explicit boundary mass terms were constructed in Refs.~\cite{manjunath2023rotational,rao2023effective,wieder2020strong}. 
It has also been argued in Refs.~\cite{may2022crystalline,manjunath2023rotational} that when rotational symmetry is unbroken and the extrinsic geometry of the boundary is ignored, boundary modes cannot be gapped when $\bar{s}\neq 0$. 
Until now, however, the role of extrinsic geometry has not been thoroughly investigated.

In what follows, we will show how the GJA term emerges from a microscopic description of the boundary theory for rotationally invariant topological phases. 
Starting from a general $K$-matrix effective field theory for a bulk abelian topological phase in 2D, we will derive the boundary action in the presence of electromagnetic and geometric perturbations.  
We will then show that when the Hall conductivity (and chiral central charge) vanishes, there generally exist rotationally invariant mass terms that couple the boundary theory to extrinsic geometry. 
The mass gaps out the boundary degrees of freedom, leaving behind a local response theory governed by the GJA action.

\section{Edge Theory for Abelian Topological Phases with Rotational Symmetry}\label{sec:kmatrix}
We will start with a generic (2+1) dimensional Abelian topological state whose long distance behavior is described by the following multi-component Abelian Chern Simons action \cite{wen1991topological},
\begin{equation}
    S_0=-\int\mathrm{d}^3X\frac{1}{4\pi}K^{IJ}\epsilon^{\mu\nu\lambda}a^I_{\mu}\partial_{\nu}a^J_{\lambda},
\end{equation}
where $K_{IJ}$ is an integer-valued symmetric matrix, which is invertible, and $a^I_{\mu}$ is an emergent gauge field. 
We will assume that the dimension of $K$ is $2N\times 2N$, with an equal number of positive and negative eigenvalues, so that the system does not have a gravitational anomaly \cite{gromov2014framinganomaly}. 
This is a necessary condition for the edge modes to be gappable~\cite{levin2013protected,callan1985anomalies}.

Now, we will couple this system to an external electromagnetic field $A_{\mu}$, and spatial curvature. 
As discussed in Sec.~\ref{sec:gjareview}, we will assume that $\mathcal{M}$ is split into space and time components, $\mathcal{M}=\mathbb{R}\times\Gamma$. 
We will not make any assumptions about the spacetime symmetries of $\mathcal{M}$. 
In other words, it is not necessary for $\mathcal{M}$ to have any rotational Killing vectors $\partial_{\theta}$ with closed orbits.
However, even on an arbitrary spatial manifold, any theory is invariant under the local internal $SO(2)\cong U(1)$ rotational symmetry Eq.~\eqref{eq:introtsym}.
We assume that there are no other bulk tensor fields with internal indices to which the fermions can couple.
Presence of tensor fields with internal indices signal a breakdown of the internal rotational symmetry, either due to  spontaneous symmetry breaking or explicit symmetry breaking (due to, e.g. a preferred lattice frame) and will be studied elsewhere. 
A theory that is invariant under $U(1)$ charge symmetry and $SO(2)\cong U(1)$ local internal rotational symmetry will have conserved currents $J_{\mu}$ and $J^s_{\mu}$ that can be written in terms of the emergent gauge fields $a^I_{\mu}$,
\begin{flalign}
    &J^{\mu}=\frac{1}{2\pi}t^I\epsilon^{\mu\nu\lambda}\partial_{\nu}a^I_{\lambda}\\
    &J_s^{\mu}=\frac{1}{2\pi}s^I\epsilon^{\mu\nu\lambda}\partial_{\nu}a^I_{\lambda},
\end{flalign}
where $t^I$ and $s^I$ are known as the charge and spin vectors respectively. 
The charges 
$t_I$ are all integers while the spins $s_I$ are integers or half integers if the system is bosonic or fermionic respectively.
Following Ref. \cite{wen1992shift}, we couple this theory to $A$ and  $\omega$ via
\begin{flalign}
\label{s(a,w)action}
    S[A,\omega]=&S_0-\int\mathrm{d}^3{X} A_{\mu}J^{\mu}-\int\mathrm{d}^3{X}\omega_{\mu}J_s^{\mu}\\
    \label{CSwA&Om}
    =&S_0-\int\frac{t^I}{2\pi} A\wedge\mathrm{d}a^I-\int\frac{s^I}{2\pi} \omega\wedge\mathrm{d}a^I.
\end{flalign}
Note that we have ignored any explicit coupling to the vielbeins $E^i_A(X)$ since we are assuming that the system is torsion free.
If $\mathcal{M}$ has a boundary $\partial\mathcal{M}=\mathbb{R}\times\Gamma$, Eq. \eqref{CSwA&Om} is not gauge invariant under transformations of $A$ and $\omega$ and we need to add a suitable boundary theory to make the action gauge and internal rotationally invariant. 
We also need a boundary condition on the emergent $a^I_{\mu}$ gauge fields to make the variational problem well defined.
We will choose for simplicity
\begin{equation}
\label{bc}
    \left. a^I_0\right|_{\partial\mathcal{M}}=v (K^{-1})^{I}_{\hphantom{I}J}\left. a^J_{x}\right|_{\partial\mathcal{M}},
\end{equation}
where $x$ is a coordinate along the boundary, and 
\begin{equation}\label{eq:bc}
    v=\frac{v_0}{\sqrt{\gamma}}
\end{equation}
is a scalar density, where $v_0$ is a scalar and $\gamma=\gamma_{00}\gamma_{xx}\equiv \gamma_{xx}$ is the determinant of the boundary induced metric.
We have assumed that the boundary metric is orthogonal in $t$ and $x$, but not necessarily time-independent.
The factor of $\sqrt{\gamma}$ in the velocity is crucial for the spatial diffeomorphism invariance of Eq. \eqref{bc}.
Indeed, under a spatial diffeomorphism, $x\rightarrow x'(x)$, $a^I_x\rightarrow a_x^I\frac{\mathrm{d}x}{\mathrm{d}x'}$, and $\gamma_{xx}\rightarrow\gamma_{xx}\big(\frac{\mathrm{d}x}{\mathrm{d}x'}\big)^2$, therefore, Eq. \eqref{bc} is invariant.
In the Appendix \ref{app:A}, we choose the more general boundary condition, $a^I_0=\frac{1}{\sqrt{\gamma}}V^I_{\hphantom{I}J}a^J_x$, where $V$ is any matrix that commute with $K$.
In other words, the $K$ and $V$ matrices are simultaneously diagonalizable, so that the propagating modes (eigenvectors of $V_{IJ})$ coincide with the anyon modes (eigenvectors of $K_{IJ}$). 
Our explicit choice in Eq.~\eqref{bc} ensures that all modes will have the same velocity $v$ for simplicity.
This generalizes the boundary condition considered in Ref. \cite{manjunath2023rotational}.
With the boundary condition Eq. \eqref{bc}, we can integrate out the emergent gauge fields $a^I_{\mu}$. 
$a^I_0$ does not appear with time derivatives in the action, so it imposes a constraint, 
\begin{equation}
    f^I_{ij}=-(K^{-1})^{IJ}t_JF_{ij}-(K^{-1})^{IJ}s_JR_{ij},
\end{equation}
where $f_{ij}^I$, $F_{ij}$ and $R_{ij}$ are the components of the curvature forms $\mathrm{d}a^I$, $\mathrm{d}A$ and $\mathrm{d}\omega$ respectively. 
This constraint can be (locally) solved by taking
\begin{equation}\label{eq:aIsol}
    a^I_i=\partial_i\phi^I-(K^{-1})^{IJ}\left(t_J A_i+s_J\omega_i\right).
\end{equation}
In order for $a^I$ to be invariant under electromagnetic and internal rotational gauge transformations, Eq.~\eqref{eq:aIsol}, requires that under $A\rightarrow A+\mathrm{d}f$,
\begin{equation}
\phi^I(X)\rightarrow\phi^I(X)+(K^{-1})^{IJ}t_Jf(X),
\end{equation}
and under a local rotation $\omega\rightarrow\omega+\mathrm{d}f$,
\begin{equation}
    \phi^I(X)\rightarrow\phi^I(X)+(K^{-1})^{IJ}s_Jf(X).
\end{equation}
Note that, by definition, the local rotations act as internal symmetries. 
This might seem to contradict the fact that in a flat (zero curvature) background, rotations act not only on the microscopic fields, but also on the underlying space-time.
However, we would like to point out that this is a peculiarity of the flat metric, $g_{ij}=\delta_{ij}$.
For a flat metric, we are allowed to work in a gauge where the frames are $e^A_i(X)=\delta^A_i$, and hence an orthogonal transformation of the internal indices $A,B,\cdots$ can be exchanged for an orthogonal transformation of the spatial indices, $i,j,\cdots$. 
In a general manifold, this identification may not be consistent, and requires a more careful treatment of diffeomorphisms.

We can now integrate out the $a^I$ gauge fields  (see Appendix \ref{app:A} for details) and see that the bulk induced action is given by
\begin{flalign}
\label{bulk}
    S_{\mathrm{bulk}}=&\frac{\sigma}{4\pi}\int_{\mathcal{M}}A\wedge \mathrm{d}A \nonumber \\&+\frac{\bar{s}}{2\pi}\int_{\mathcal{M}}A\wedge \mathrm{d}\omega+\frac{l_s}{4\pi}\int_{\mathcal{M}}\omega\wedge \mathrm{d}\omega,
\end{flalign}
where 
\begin{equation}
\sigma=t^TK^{-1}t,\;\; \bar{s}=s^TK^{-1}t,\;\; l_s=s^TK^{-1}s.
\end{equation}
Note that our assumption that $K$ has an equal number of positive and negative eigenvalues implies that the chiral central charge is zero, hence the framing anomaly term does not appear, \cite{gromov2014framinganomaly}.

Using the boundary condition Eq. \eqref{bc}, the boundary terms that arise when we integrate out the $a^I$ combine to yield the boundary action (see appendix \ref{app:A} for a derivation) 
\begin{widetext}
\begin{flalign}
\label{bndy}
    S_{\mathrm{bndy}}=-\int_{\partial\mathcal{M}}\mathrm{d}t\mathrm{d}x\Big[&\frac{K^{IJ}}{4\pi}\partial_x\phi_I\partial_t\phi_J-\frac{v}{4\pi}(\partial_x\phi^I)^2 
    -\frac{\partial_x\phi^I}{2\pi}\left(t_IA_t-vq_IA_x\right)-\frac{\partial_x\phi^I}{2\pi}\left(s_I\omega_t-vl_I\omega_x\right) \Big. \nonumber \\ &+\frac{\sigma}{4\pi}A_tA_x-v\frac{\sigma'}{4\pi}A_x^2+\frac{\bar{s}}{2\pi}\omega_xA_t-v\frac{s'}{2\pi}\omega_xA_x+\frac{l_s}{4\pi}\omega_t\omega_x-v\frac{l_s'}{4\pi}\omega_x^2\Big],
\end{flalign}
\end{widetext}
where 
\begin{align}
q_I&=(K^{-1})^{IJ}t_J,\;\; l_I=(K^{-1})^{IJ}s_J,\;\; \sigma'=q^Iq_I, \\
s'&=q^Il_I,\;\; l_s'=l^Il_I,
\end{align} 
and $v$ depends on the metric via Eq.~\eqref{eq:bc}. 
The dependence of $v$ on $\gamma$ ensures that the kinetic Lagrangian for the bosons is a scalar. 
We would like to emphasize the importance of the $\phi$-independent counter-terms in Eq. \eqref{bndy}; without them the combined theory $S_\mathrm{bulk}+S_\mathrm{bndy}$ will not be gauge invariant. 

Our primary goal is to now determine when the boundary modes $\phi^I$ can be gapped without breaking gauge or internal rotational symmetries. 
When the Hall conductivity $\sigma\neq 0$, a gapped boundary is impossible since the boundary bosons have a gauge anomaly, and therefore, must be gapless due to anomaly inflow from the bulk. 
We will thus assume that $\sigma=0$ going forward.
When $\sigma=0$ the bulk theory is local internal rotationally invariant but not gauge invariant.
Since the bulk theory is not gauge invariant, the boundary theory needs to be anomalous to conserve charge.
Under a gauge transformation and internal rotation $A\rightarrow A+\mathrm{d}f$, $\omega\rightarrow \omega+\mathrm{d}g$, the bulk action changes according to,
\begin{equation}
\label{anomaly}
    \Delta S_{\mathrm{bulk}}=-\frac{\bar{s}}{2\pi}\int_{\partial\mathcal{M}}f \mathrm{d}\omega-\frac{l_s}{4\pi}\int_{\partial\mathcal{M}}g \mathrm{d}\omega.
\end{equation}
The minus signs arise due to the spacetime orientation of $\partial\mathcal{M}$, following the conventions of Ref. \cite{gromov2016boundary}
This is exactly compensated by the (anomalous) change in the boundary action where the bosons transform according to,
\begin{equation}
\label{eq:phigaugetrans}
    \phi^I\rightarrow\phi^I+q^If+l^Ig.
\end{equation}
As we will show in the next section, provided the $K$ matrix satisfies certain conditions (that are independent of the internal rotational symmetry), there exist internal rotationally invariant mass terms on the boundary that (with $\sigma=0$) can gap out the boundary modes.
Therefore, the change in the bulk action, Eq. \eqref{anomaly}, will actually be compensated by the local Lagrangian that lives on the boundary, which we obtain by integrating out the gapped boundary bosons, as predicted in Ref. \cite{gromov2016boundary}. 

\section{Effective action for the bulk bosons}
\label{sec:effection_action}

We want to add mass terms to the action Eq. \eqref{bndy} that gaps out all the boundary modes which can then be integrated out. 
The conditions required for this to be possible were given, in the absence of symmetry, in Ref.~\cite{levin2013protected}. 
There it was shown that the $\phi^I$ modes can be gapped out if and only if there exists a Lagrangian subgroup $\mathcal{L}$ of integer vectors $m_i$ such that:
\begin{itemize}
    \item $m^TK^{-1}m'$ is an integer for all $m,m'\in\mathcal{L}$
    \item If $l$ is not ``equivalent'' to any element of $\mathcal{L}$ (i.e. $l\neq m_i+K\Lambda$ for some integer vector $\Lambda$), then $m^TK^{-1}l$ is not an integer for some $m\in\mathcal{L}$.
\end{itemize}
If these conditions are satisfied, then the $\phi_I$ fields can be gapped out by backscattering terms of the form,
\begin{equation}
\label{massterm}
    U(\Lambda_i)=U_i(x)\cos\Big(\Lambda^T_iK\phi(x)-f_i(x)\Big),
\end{equation}
where $\Lambda_i$ is a set of linearly independent integer vectors with $i\in\{1,\cdots,N\}$ satisfying the null vector condition $\Lambda_i^TK\Lambda_j=0$, and $f_i(x)$ is any smooth function. 
It was pointed out in Ref. \cite{levin2013protected} that the existence of the set of $\Lambda_i$ is equivalent to the conditions listed above.
We can get a sense for the importance of the null-vector condition as follows:\cite{PhysRevB.86.125119}
Canonical quantization of the action Eq.~\eqref{bndy} show that the bosons obey the Kac-Moody algebra
\begin{equation}
    [\Lambda_i^T\partial_x\phi(x),\Lambda_j^T\partial_y\phi(y)]=2\pi i(\Lambda_i^TK^{-1}\Lambda_j)\delta'(x-y),
\end{equation}
or equivalently,
\begin{equation}
    [\Lambda_i^TK\partial_x\phi(x),\Lambda_j^TK\partial_y\phi(y)]=2\pi i(\Lambda_i^TK\Lambda_j)\delta'(x-y),
\end{equation}
where $\delta'(x)=\partial_x\delta(x)$.
Therefore, if $\Lambda_i^TK\Lambda_j=0$, the mass terms are all mutually commuting at every point, and so can be pinned at certain classical values simultaneously, gapping out the edge excitations.
Because our bosons are charged under electromagnetic gauge transformations, we need to augment the Lagrangian subgroup conditions with the additional neutrality constraints $\Lambda_i^Tt=0$ for all $i$. 
This additional constraint ensures that the mass term Eq. \eqref{massterm} is invariant under local gauge transformations, $\phi^I(x)\rightarrow\phi^I(x)+q^Ig(x)$,
\begin{align}
    U(\Lambda_i)&\rightarrow U_i(x)\cos\Big(\Lambda^T_iK\phi(x)+(\Lambda_i^Tt)g(x)-f_i(x)\Big)\nonumber \\
    & = U(\Lambda_i).
\end{align}
As pointed out in Ref. \cite{levin2013protected}, the Lagrangian subgroup condition along with the additional neutrality constraints $\Lambda_i^Tt=0$ for all $i$, forces $\sigma_{xy}=0$. 
To see this, note that since $\Lambda_i$ is orthogonal to $K\Lambda_i$s, and the set of $\{K\Lambda_i\}$ are linearly independent, we must have that, 
\begin{equation}
t=\sum_{i=1}^Na_iK\Lambda_i
\end{equation}
and hence
\begin{equation}
\sigma_{xy}=t^TK^{-1}t=\sum_{i=1}^Na_it^T\Lambda_i=0.
\end{equation}
We will now show that rotational invariance does not place any further constraints on the mass terms. 

Assume that we have a complete set of $\Lambda_i$'s satisfying the null vector and neutrality constraints,
\begin{align}
\Lambda_i^T K \Lambda_j &= 0 \\
\Lambda^T_it&=0
\end{align}
ignoring rotational invariance. 
On the boundary, the spin connection $\omega$ is related to the extrinsic curvature according to Eq.~\eqref{eq:alphadef},
\begin{equation}
    \omega|_{\partial\mathcal{M}}+K=\mathrm{d}\alpha,
\end{equation}
where $\alpha(x)$ measures the change in angle between the bulk frames parallel transported around the boundary, measured relative to the boundary normal and tangent vectors.
Under a local rotation, $\omega\rightarrow\omega+\mathrm{d}f$, $\alpha\rightarrow\alpha+f$.
Therefore, the linear combinations
\begin{equation}\label{eq:tildefielddef}
    \tilde{\phi}^I(x)=\phi^I(x)-l^I\alpha(x)
\end{equation}
are internal rotationally invariant. 
Using this, we see that the mass term Eq. \eqref{massterm} can be made rotationally invariant if we choose $f_i(x)=(\Lambda_i)_{I}s^I\alpha(x)$.
In other words, we can write rotationally invariant mass terms
\begin{equation}
\label{multi_boson_mass}
U(\Lambda_i)=U_i(x)\cos\Big[\Lambda_i^TK\Big(\phi(x)-l\alpha(x)\Big)\Big].
\end{equation}
This is one of the  central results of this paper and shows that rotational invariance does not place any constraints on the Wen-Zee shift, $\bar{s}$.
Note that $\alpha(x)$ is not periodic on manifolds with a non-trivial Euler characteristic, $\chi(\Gamma)$, as,
\begin{equation}
\label{alpha_period}
    \int_{\Gamma}\mathrm{d}\omega+\int_{\partial\Gamma}K=\int_{\partial\Gamma}\mathrm{d}\alpha=2\pi\chi(\Gamma).
\end{equation}
Eq. \eqref{alpha_period} implies $\alpha(x+L)=\alpha(x)+2\pi\chi(\Gamma)$, with integer $\chi(\Gamma)$ due to the Gauss-Bonnet theorem.
Therefore, we need to make sure the mass term, Eq. \eqref{multi_boson_mass} is periodic, when $s^I$ has both integer and half-integer components. 
To do so, we will carefully examine the boundary conditions on the boson fields via canonical quantization.

First, following Refs. \cite{wang_2015_boundary,wang_2023_nonperturbative}, we expand the bosons as,
\begin{equation}\label{eq:modeexpansion}
    \phi^I(x)=Q^I+(K^{-1})^I_{\hphantom{I}J}P^J\frac{2\pi}{L}x+i\sum_{n\neq 0}\frac{1}{n}a^I_ne^{-i\frac{2\pi nx}{L}},
\end{equation}
where $Q^I$ is the center of mass of the boson (the zero mode) and $P^I$ is the momentum of the center of mass (the winding).
From the commutation relation between the canonical momentum, $\Pi^I=\frac{1}{2\pi}K^{IJ}\partial_x\phi_J$, and $\phi^I(x)$, given by $[\phi^I(x),\Pi^J(y)]=i\delta^{IJ}\delta(x-y)$, we deduce that the winding modes $Q^I$, $P^J$ satisfy $[Q^I,P^J]=i\delta^{IJ}$ and the oscillator modes satisfy, $[a^I_n,a^J_m]=n(K^{-1})^{IJ}\delta_{n,-m}$. 
Since our bosonic fields arise as the phase of a $U(1)$ gauge transformation via Eq.~\eqref{eq:aIsol}, we must identify $\phi^I$ with $\phi^I+2\pi$ (i.e. the bosons are compact with compactification radius $R=1$), which enforces that the zero modes $Q$ and $Q+2\pi$ are equivalent. 
This restricts $P^I$ to have integer eigenvalues.

The mass term Eq. \eqref{multi_boson_mass}, is the real part of the following vertex operator,
\begin{equation}
 \label{vertex_op}   
 V_i(x)=\exp\Big[i\Lambda_i^T\Big(K\phi(x)-s\alpha(x)\Big)\Big],
\end{equation}
which has the net zero mode part,
\begin{flalign}
\label{vertex_op_2}
    &\exp\Big[i\Lambda_i^T\Big(KQ+i\Lambda_i^TP\frac{2\pi}{L}x-s\alpha(x)\Big)\Big]\nonumber\\&=\exp \Big[i\Lambda_i^TKQ\Big]\exp\Big[i\Lambda_i^TP\Big(\frac{2\pi}{L}x\Big)\Big]e^{-i\Lambda_i^Ts\alpha(x)},
\end{flalign}
where we have used the null vector condition and the Baker-Campbell-Hausdorff formula to reorder the exponentials.
Since $\alpha(x+L)=\alpha(x)+2\pi\chi(\Gamma)$, for integer $s^I$ (which is the case for bosonic systems), Eq. \eqref{vertex_op} is a periodic function of $x$ since $P^I$ has integer eigenvalues. 
If $s^I$ has a half integer component (which is possible for fermionic systems), we can exploit the fact that the null vectors $\Lambda_i$ span an integer lattice to choose instead for our mass terms
\begin{equation}
        U(\Lambda_i)=U_i(x)\cos\Big[2\Lambda_i^TK\Big(\phi(x)-l\alpha(x)\Big)\Big],
\end{equation}
which is periodic. 
For convenience, let us define
\begin{equation}
    \tilde{\Lambda}_i=\begin{cases}
        \Lambda_i,\;\; s\in \mathbb{Z}^{2N} \\
         2\Lambda_i,\;\; s\in (\mathbb{Z}+\frac{1}{2})^{2N},
    \end{cases}
\end{equation}
such that $U(\tilde{\Lambda}_i)$ is always periodic.

We can now isolate the degrees of freedom of the system in the basis of the $\Lambda_i$'s in order to explicitly integrate out the boundary modes in the presence of the mass terms. 
We first decompose the $\phi_I$ bosons according to
\begin{flalign}
    \label{decom}
    \phi^I(x)&=\sum_{i=1}^Nc_i(x)\Lambda_i^I+\sum_{i=1}^{N}b_i(x)(K\Lambda_i)^I \nonumber \\
    &\equiv\eta^I(x)+\beta^I(x),
\end{flalign}
which is allowed since $\Lambda_i^TK\Lambda_j=0$ for all $i$ and $j$, and we are assuming we have a complete set of $\Lambda_i$'s. 
We also separate $s_I$, $l_I$, $t_I$ and $q_I$ into components,
\begin{flalign}
    l^I&=\sum_{i=1}^Np_i\Lambda_i^I+q_i(K\Lambda_i)^I\equiv l_1^I+l_2^I\\
    s^I&=\sum_{i=1}^Nm_i\Lambda_i^I+n_i(K\Lambda_i)^I\equiv s_1^I+s_2^I \label{eq:sdecomp}\\
    t^I&=\sum_{i=1}^Na_iK\Lambda_i\\
    q^I&=K^{-1}t=\sum_{i=1}^Na_i\Lambda_i,
\end{flalign}
where we have used the orthogonality of $t^I$ and $\Lambda_i$ for all $i$. 
Under local rotations, $\eta\rightarrow\eta+l_1 f$ and $\beta\rightarrow\beta+l_2 f$. 
Under electromagnetic gauge transformations, $\beta$ does not transform while $\eta\rightarrow \eta+qf$
$s_1$, $s_2$, $l_1$ and $l_2$ have rational components. 
Following \cite{levin2013protected}, we choose a basis such that $\Lambda_i^T=(v_i,0)$, where $v_i$ is an $N$-component vector. 
In this basis
 $K$ takes the form
\begin{equation}
\label{Kmat}
    K=\begin{pmatrix}
        0&A\\
        A^T&B
    \end{pmatrix},
\end{equation}
where $A$ and $B$ are $N\times N$ matrices, and $A$ is invertible because $K$ is invertible.
The first block of $K$  is zero due to the null vector condition $\Lambda_i^TK\Lambda_j=0$.
Furthermore, $B$ is a symmetric matrix since $K$ is symmetric.
The inverse of $K$ is
\begin{equation}
    K^{-1}=\begin{pmatrix}
        -(A^T)^{-1}BA^{-1}&(A^T)^{-1}\\
        A^{-1}&0
    \end{pmatrix}.
\end{equation}
From the relation $Kl=s$, we can deduce that $Al_2=s_1$ and $A^Tl_1+Bl_2=s_2$.
From $Kq=t$, we also get $A^Tq=t$

In terms of the $\eta^I$ and $\beta^I$ modes, the mass term, Eq. \eqref{multi_boson_mass} becomes, upon inserting Eq.~\eqref{Kmat}, \eqref{eq:sdecomp} and \eqref{decom},
\begin{equation}\label{eq:massintermsofbeta}
    U(\tilde{\Lambda}_i)=U_i(x)\cos\Big(\tilde{\Lambda}_i^TA\beta(x)-\tilde{\Lambda}_i^Ts_1\alpha(x)\Big),
\end{equation}
We see that only the $\beta$ modes appear in the mass terms. 
From the action Eq. \eqref{eta,betaeq} in Appendix \ref{app:B}, we see that the $\eta^I$ modes do not appear with time derivatives, and are canonically conjugate variables to $A^{IJ}\beta_J$. 
Therefore, they are not genuine degrees of freedom and can be integrated out to get a local action for only the $\beta^I$ modes.
Exploiting the fact that the $\eta^I$ modes appear quadratically in the action, we integrate them out by using their classical equations of motion (this is done in Appendix \ref{app:B}).
Once we do that, the action for the $\beta^I(x)$s becomes,
\begin{widetext}
\begin{flalign}
\label{gapped}
    S_{\mathrm{bndy}}=-\int_{\partial\mathcal{M}}&\mathrm{d}t\mathrm{d}x\Big[\frac{B^{IJ}}{4\pi}\partial_x\beta_I\partial_t\beta_J-\frac{v}{4\pi}\Big(\partial_x\beta^I-l^I_2\omega_x\Big)^2+\frac{1}{4\pi v}\Big(A^{IJ}\partial_t\beta_J-s_1^I\omega_t\Big)^2-\frac{t^I}{2\pi}\Big(A_t\partial_x\beta^I-A_x\partial_t\beta^I\Big)\Big.\nonumber\\&-\frac{s_2^I}{2\pi}\Big(\omega_t\partial_x\beta^I-\omega_x\partial_t\beta^I\Big)-\frac{1}{2\pi}B^{IJ}l^J_2\omega_x\partial_t\beta^I-\frac{\bar{s}}{2\pi}\omega_tA_x+\frac{\bar{s}}{2\pi}\omega_xA_t+
    \frac{l_s}{4\pi}\omega_t\omega_x-\frac{1}{2\pi}s^1_Il^1_I\omega_x\omega_t+\mathcal{L}_{\mathrm{mass}}\Big]
\end{flalign}
\end{widetext}
where 
\begin{equation}
\mathcal{L}_{\mathrm{mass}}=\sqrt{\gamma}\sum_i U(\tilde{\Lambda}_i)
\end{equation}
contains the gapping terms \eqref{massterm}.
We would like to point out that integrating out $\eta^I$ is not possible when $\sigma\neq 0$. 
This is because, when the Hall conductance is non-zero, $\eta^I$ and $A^{IJ}\beta_J$ are not canonically conjugate variables anymore.
There is an additional term in the kinetic part of the action, which is quadratic in the $\eta^I$, so the $\eta^I$s become genuine degrees of freedom.

Note that our mass terms in Eqs.~\eqref{eq:massintermsofbeta} circumvent the no-go argument of Ref.~\cite{manjunath2023rotational} due to the explicit coupling to extrinsic geometry $\alpha$. 
This highlights the fact that geometric gauge fields can behave differently to gauge fields associated to purely internal symmetries that know nothing about the underlying geometry. 
We additionally note that the flux insertion arguments of Ref.~\cite{manjunath2023rotational} cannot be directly applied to our edge theory, since internal rotational symmetry fluxes change the ambient geometry.

Finally, if the mass terms $|U_i(x)|$ are large, we can explicitly integrate out the $\beta_I$ modes in the saddle point approximation. 
This is equivalent to pinning the fields $A^{IJ}\beta_J(x)$ to their average value to minimize the energy due to the mass terms, 
\begin{equation}\label{eq:saddle}
    A^{IJ}\beta_J(x)\rightarrow \langle A^{IJ}\beta_J(x)\rangle = s_1^U\alpha(x).
\end{equation}
The null vector condition on $\Lambda_i$ ensures that we can perform the saddle point approximation independently for each mass term without violating the canonical commutation relations. 
In terms of our original bosonic fields $\phi_I$, Eq.~\eqref{eq:saddle} corresponds to setting
\begin{equation}\label{eq:saddleoriginal}
    K^I_{\hphantom{I}J}\langle\phi^J\rangle\sim s^I\alpha(x).
\end{equation}
At first glance, this seems potentially problematic when $s_I$ has half-integer components since, via Eq.~\eqref{eq:modeexpansion} $\phi^I(x)$ satisfies, $\phi^I(x+L)=\phi^I(x)+2\pi(K^{-1})^I_{\hphantom{I}J}P^J$, with integer eigenvalues for $P^I$.
However, the saddle point ground state need not be an eigenstate of $P_I$.
Note that using the saddle point condition Eq.~\eqref{eq:saddleoriginal}, the  vertex operators corresponding to the trivial degrees of freedom still have the right statistics.
If $n_IK_{IJ}$ is odd for some vector $n_I$ of integers, then vertex operator,
\begin{equation}
    V^n(x)=\exp\Big[in_IK^I_{\hphantom{I}J}\phi^J(x)\Big],
\end{equation}
corresponds to a fermion, and is antiperiodic in $x$ as an operator.
$V^n(x)$ is antiperiodic even after we use the mass term to pin the value of the boson $\phi^I(x)$, since, $   V^n(x)\sim\exp\Big[in_IK^I_{\hphantom{I}J}\langle\phi^J\rangle\Big]=e^{in_Is^I\alpha(x)}$
is antiperiodic for half-integer $n_Is^I$ as required.

We can now use the saddle point condition Eq.~\eqref{eq:saddle} to integrate out the $\beta_I$ modes in the boundary action Eq.~\eqref{gapped}. 
Doing so, we find that the boundary theory reduces to,
\begin{flalign}
    \label{GJA}
    S_{\mathrm{bndy}}=&\int \mathrm{d}t\mathrm{d}x\Big[\frac{\bar{s}}{2\pi}\Big(K_xA_t-K_tA_x\Big)+\frac{l_s}{4\pi}\Big(K_x\omega_t-K_t\omega_x\Big)\Big.\nonumber\\&-\frac{s_1^Ts_1}{4\pi v}K_t^2+v\frac{l_2^Tl_2}{4\pi }K_x^2+\frac{l_1^Ts_1-l_2^Ts_2}{4\pi}K_xK_t\Big],
\end{flalign}
where $K_{\bar{\mu}}$ is the extrinsic curvature of the boundary.
In Appendix \ref{app:B} we briefly discuss how the action is modified when $v=0$.
We see that the first line in Eq.~\eqref{GJA} is the Gromov-Jensen-Abanov action Eq.~\eqref{eq:gjaintro} first introduced in Ref.~\cite{gromov2016boundary}. 
The remaining terms in the boundary theory represent the local elastic energy density of the boundary.

We have thus shown that rotational symmetry does not place any new constraints on the stability of the edge theory for abelian topological phases, due to the coupling to external geometry. 
In demonstrating this, we have also shown how the GJA action arises as the induced action of the gapped boundary. 
Up to now, we have formulated our arguments in terms of the effective $K$-matrix Chern-Simons theory for the bulk. 
Now, we will demonstrate how our arguments apply in a concrete microscopic model.

\section{Microscopic Example: Rotationally Invariant Higher-Order Topological Insulators}\label{sec:example}

As a concrete example, we will now re-examine the microscopic theory for a rotationally invariant $\mathbf{k}\cdot\mathbf{p}$ theory of a higher-order topological insulator first presented in Ref.~\cite{may2022crystalline}. 
The microscopic model we consider has the following bulk Lagrangian in a flat background ($g_{ij}=\delta_{ij}$),
\begin{equation}
    \label{hoti lagrangian}
    \mathcal{L}=\bar{\Psi}(i\gamma^0\partial_t+iv\gamma^{i}\partial_{i}+m\tau_3)\Psi,
\end{equation}
where $\Psi$ is a 4-component Dirac spinor and the three gamma matrices and mass matrix are, respectively,
\begin{align}
\gamma^0&=\sigma^0\otimes\sigma^3,\;\; \gamma^1=i\sigma^3\otimes\sigma^1,\;\; \\
\gamma^2&=i\sigma^3\otimes\sigma^2\;\; \tau^3=-\sigma^3\otimes\sigma^0.
\end{align}
The corresponding Hamiltonian density is,
\begin{equation}
    \label{k.p ham}
    H(\mathbf{X})=\sigma^3\otimes\Big[m\sigma^3-iv\Big(\sigma_1\frac{\partial}{\partial X_2}-\sigma_2\frac{\partial}{\partial X_1}\Big)\Big].
\end{equation}

In a flat background, this theory has the following symmetries, along with translation invariance:
\begin{itemize}
    \item Gauge symmetry, which acts internally,
    \begin{equation}
        \Psi(X)\rightarrow\Psi'(X)=e^{i\xi}\Psi(X).
    \end{equation}
    To gauge this symmetry, we introduce the electromagnetic gauge field $A_{\mu}$ and modify the partial derivative to $D_{\mu}=\partial_{\mu}-iA_{\mu}$.
    \item Rotational symmetry, which acts as a space-time symmetry,
    \begin{equation}
        \Psi(X)\rightarrow\Psi'(X')=e^{i\xi\frac{\gamma^0}{2}}\Psi(X),
    \end{equation}
    where $X'=(t,R(\xi)\mathbf{X})$ with $R(\xi)$ a $2\times 2$ rotation matrix. 
    In order to gauge this symmetry, we first make it an internal symmetry by introducing the vielbeins $E^i_A(X)$ and the non-relativistic spin connection $\omega_{\mu}$.
    Then we modify the covariant derivative to $E^i_AD_i=E^i_A(X)\Big(\partial_i-iA_i-i\omega_i\frac{\gamma^0}{2}\Big)$. 
    \item Isospin symmetry, which acts internally,
    \begin{equation}
        \Psi(X)\rightarrow\Psi'(X)=e^{i\xi\tau_3}\Psi(X).
    \end{equation}
    We gauge this symmetry by adding another $U(1)$ gauge field, $\tilde{A}_{\mu}$ and modify the covariant derivative to, $E^i_
    AD_i=E^i_A(X)\Big(\partial_i-iA_i-i\omega_i\frac{\gamma^0}{2}-i\tilde{A}_i\tau_3\Big)$.
\end{itemize}
Hence, in a curved manifold, the locally symmetric Lagrangian for the $\mathbf{k}\cdot\mathbf{p}$ Hamiltonian \eqref{hoti lagrangian} is,
\begin{equation}
\label{curvedaction}
    \mathcal{L}=\sqrt{g}\bar{\Psi}(i\gamma^0D_t+iv\gamma^AE^i_A(X)D_i+m\tau_3)\Psi,
\end{equation}
with $D_{\mu}=\partial_{\mu}-iA_{\mu}-i\omega_{\mu}\frac{\gamma^0}{2}-i\tilde{A}_{\mu}\tau_3$ and we have chosen the nonrelativistic gauge $E^t_0=1$.
We will restrict our attention in this section to torsion-free geometries that do not change with time. 
Hence we can choose our vielbeins to be time-independent, and the spin connection will only have space components.
Following refs. \cite{manjunath2021crystalline,zhang2022fractional,barkeshli2025disclinations}, we will identify the local isospin gauge field with the spin connection of the fermions, $\tilde{A}_{\mu}=\omega_{\mu}$, as Eq. \eqref{curvedaction} is the low-energy limit of a tight-binding model that only has the combined fourfold rotational symmetry \cite{may2022crystalline, benalcazar2017electric}, $\Psi\rightarrow\exp i\frac{\pi}{2}\Big(\frac{\gamma^0}{2}+\tau_3\Big)\Psi$\footnote{We would like to point out that on arbitrary manifolds, this identification may not be valid. 
More generally we could have $\tilde{A}_{\mu}=\tilde{\omega}_{\mu}$, where $\tilde{\omega}$ is a discrete lattice ``anisospin'' connection~\cite{gromov2017investigating}, that is determined by how the lattice is embedded inside the manifold.}. 
The continuous rotational symmetry in the low energy limit is an emergent symmetry.

The bulk response of this theory was analyzed in Ref. \cite{may2022crystalline} and the authors showed that the response is given by the Wen-Zee action,
\begin{equation}
\label{wenzee}
    \mathcal{L}_{\mathrm{geom}}=\frac{1-\mathrm{sgn}(m)}{2\pi}\epsilon^{\mu\nu\sigma}A_{\mu}\partial_{\nu}\omega_{\sigma},
\end{equation}
where they also identified the isospin symmetry of the system with the local internal rotational symmetry.

In order to analyze the edge modes of this model, we consider a domain wall between the trivial and non-trivial phases.
If $\mathbf{X}=(X^1,X^2)$ are coordinates on the manifold, the curved domain wall has two representations, as a level curve of a function, $F(\mathbf{X}) = $ constant, or the parametric representation, $(X^1(\theta), X^2(\theta))$
where $\theta \sim \theta + 2\pi $.
We will assume that a neighborhood of the domain wall can be foliated by non-intersecting curves $F(\mathbf{X})=$ constant, and we install the same coordinate $\theta$ on each curve.
The normal one-form to these curves is given by $N=\mu\mathrm{d}F$, for some function $\mu$ such that $g^{ij}N_iN_j=1$.
Following \cite{Weinberg:1972kfs}, we construct a Gaussian coordinate system in a neighborhood of the domain wall by considering the flow of the vector field $N_i$ through $X^i(\theta)$, defined by the equation,
\begin{equation}
\label{flow}
    \frac{\partial X^i(\rho,\theta)}{\partial\rho}=N^i(\mathbf{X}(\rho,\theta))=g^{ij}N_j,
\end{equation}
with the initial condition $X^i(R,\theta)=X^i(\theta)$.
This places the domain wall at $\rho=R$.
Eq. \eqref{flow} is a generalization of the near-boundary coordinates considered in Ref. \cite{dacosta1981quantum} to curved manifolds.
We restrict to geometries where $(\rho,\theta)$ can be made an orthogonal system of coordinates. 
This requires the flow, Eq.~\eqref{flow} to be orthogonal to the curves $F(\mathbf{X})=$ constant, which are the flows generated by the tangent vectors $T^i=\frac{\partial X^i}{\partial\theta}$.
The two coordinates $\theta$ and $\rho$ are orthogonal when the Lie derivative of $T$ along $N$ vanishes,
\begin{equation}\label{eq:lie}
    \mathcal{L}_NT=N^i\partial_iT^j-T^i\partial_iN^j=0.
\end{equation}
Using the flow equation Eq. \eqref{flow}, and the definition of $T^i=\frac{\partial X^i}{\partial\theta}$, the Lie derivative reduces to
\begin{equation}
\mathcal{L}_NT=\frac{\partial^2 X^j}{\partial\rho\partial{\theta}}-\frac{\partial^2 X^j}{\partial\theta\partial{\rho}}=0.
\end{equation}
From the relation $T^iN_i=0$, we have $\frac{\partial F}{\partial\theta}=0$ and from $N^iN_i=1$, we have $\mu\frac{\partial F}{\partial\rho}=1$.
We deduce that in the $(\rho,\theta)$ coordinates, $N=\mathrm{d}\rho$. 
We will restrict our attention to geometries where Eq.~\eqref{eq:lie} holds; this will be necessary for us to explicitly solve for the boundary theory.

Putting this all together, we can write the bulk spatial line element, $\mathrm{d}l^2$, near the domain wall as,
\begin{flalign}
\label{gauge}
    \mathrm{d}l^2&=g_{ij}\frac{\partial X^i}{\partial \rho}\frac{\partial X^j}{\partial \rho}\mathrm{d}\rho^2\nonumber\\&+2g_{ij}\frac{\partial X^i}{\partial \rho}\frac{\partial X^j}{\partial \theta}\mathrm{d}\rho\mathrm{d}\theta
    +g_{ij}\frac{\partial X^i}{\partial \theta}\frac{\partial X^j}{\partial \theta}\mathrm{d}\theta^2\nonumber\\&=\mathrm{d}\rho^2+q^2(\rho,\theta)\mathrm{d}\theta^2.
\end{flalign}
When written in terms of the $(\rho,\theta)$ coordinates, the metric restricted to the boundary gives the induced metric on the boundary, which is,
\begin{equation}
    \mathrm{d}l^2\Big|_{\partial\mathcal{M}}=q^2(R,\theta)\mathrm{d}\theta^2.
\end{equation}
There is a canonical choice of vielbeins for the metric \eqref{gauge} near the domain wall, given by, $\bar{e}^1=\mathrm{d}\rho$ and $\bar{e}^2=q\mathrm{d}\theta$, 
such that $(\bar{e}^1)^2+(\bar{e}^2)^2=\mathrm{d}s^2$. 
$\bar{e}^1$ and $\bar{e}^2$ coincide, up to a scale factor, to the boundary normal and tangent vectors.
Any other choice of vielbein will be related to this one by,
\begin{equation}
    e^A_i(\phi)=R^A_{\hphantom{A}B}(\phi)\bar{e}^B_i.
\end{equation}
The spin connection in this gauge is, $\bar{\omega}=-\partial_{\rho}q\mathrm{d}\theta$.
The extrinsic curvature of the boundary is $K_{\theta}=-\bar{\omega}_{\theta}=\partial_{\rho}q|_{\rho=R}$. 
Note that the $\phi=0$ gauge cannot be extended over the whole of the bulk manifold, since the Gauss-Bonnet theorem Eqs.~\eqref{eq:gaussbonnet} and \eqref{eq:alphadef} must hold. 
Nevertheless, near the boundary the $\phi=0$ gauge is particularly illuminating as it connects the intrinsic and extrinsic geometry of the boundary. 
In the remainder of this section, we will use the $\phi=0$ gauge to study the zero energy Jackiw-Rebbi zero energy modes of Lagrangian Eq. \eqref{curvedaction}.
These modes are exponentially localized near the boundary, and therefore we are allowed to choose the $\phi=0$ gauge because in a small enough neighborhood of the boundary, this gauge is well-defined.

Next, following Ref. \cite{Aoki_2022_curved_domain}, we analyze the curved domain wall zero modes by letting the mass term in Eq.~\eqref{k.p ham} depend on the radial coordinate, $m\rightarrow m(\rho)$. 
We first choose a $\rho_*<R$ inside the domain where the coordinates Eq. \eqref{gauge} are still valid.
We choose $m(\rho)=-|M_0|$ for $\rho<\rho_*$.
Similarly, we pick a $\rho^*>R$ and let $m(\rho)=|M_0|$, for $\rho>\rho^*$.
Finally, we let $m(\rho)$ smoothly vary from $-|M_0|$ to $|M_0|$ for $\rho_*<\rho<\rho^*$.
Since the vielbeins do not explicitly depend on time, we can write down the Hamiltonian corresponding to Lagrangian, Eq. \eqref{curvedaction},

\begin{flalign}
\label{radial_Hamiltonian_in_arb_gauge}
H(\phi)=-&\Big[m(\rho)\gamma^0\tau_3+iv\gamma^0\gamma^AE^i_A(\phi)\partial_i\nonumber\\&+v\gamma^0\gamma^AE^i_A(\phi)\omega_i\Big(\frac{\gamma^0}{2}+\tau_3\Big)\Big],
\end{flalign}
where we have set the electromagnetic gauge field to zero and used the coordinates \eqref{gauge}.
In the gauge $\phi=0$, the Hamiltonian is particularly simple,
\begin{flalign}
\label{rad ham}
    H(0)&=\sigma^3\otimes\Big[m(\rho)-iv\sigma^2\partial_{\rho}+i\frac{v}{q}\sigma^1\partial_{\theta}\Big]\nonumber\\&+v\frac{\bar{\omega}_{\theta}}{q}\Big[\frac{i}{2}\sigma^3\otimes\sigma^2+\sigma^0\otimes\sigma^1\Big].
\end{flalign}
We will now add a mass term to the Hamiltonian, Eq. \eqref{radial_Hamiltonian_in_arb_gauge}, that respects the local rotations, symmetry,
\begin{equation}
\label{bulkmass}    
U_{\phi}(\rho,\theta)=MV(\rho)\Big(\cos2\phi\gamma^0\tau_1+\sin2\phi\gamma^0\tau_2\Big),
\end{equation}
where $\tau_1=-\sigma^2\otimes\sigma^3$ and $\tau_2=-\sigma^1\otimes\sigma^3$.
Here, $V(\rho)$ is a potential that falls off exponentially with distance from the boundary at $\rho=R$.
We can write the mass term using the boundary normal and tangent vectors by using,
\begin{equation}
    2\cos\phi=N^ie^1_i+\tilde{T}^ie^2_i,
\end{equation}
and,
\begin{equation}
    2\sin\phi=N^ie^2_i-\tilde{T}^ie^1_i,
\end{equation}
with $\tilde{T}=\frac{1}{q}T$, has been rescaled so that it is normalized.
We used the normal and tangent vectors (extended to a neighborhood of the boundary) instead of $\bar{e}^1$ and $\bar{e}^2$ to make it explicit that they do not transform under internal rotations. 
We then see that the mass Eq.~\eqref{bulkmass} explicitly couples the microscopic degrees of freedom to the extrinsic geometry of the boundary.

Let us verify that Eq.~\eqref{bulkmass} is indeed rotationally invariant. 
Under an internal rotation, $e^A_i\rightarrow R^A_{\hphantom{A}B}(f)e^B_i$, we see that $\phi\rightarrow\phi+f$.
Next, using
\begin{equation}
    e^{-if\tau_3}\gamma^0\tau_1e^{if\tau_3}=(\cos2f)\gamma^0\tau_1-(\sin2f)\gamma^0\tau_2,
\end{equation}
and,
\begin{equation}
    e^{-if\tau_3}\gamma^0\tau_2e^{if\tau_3}=(\cos2f)\gamma^0\tau_2+(\sin2f)\gamma^0\tau_1,
\end{equation}
we see that,
\begin{equation}
    U_{\phi-f}(\rho,\theta)=e^{-if\Big(\frac{\gamma^0}{2}+\tau_3\Big)}U_{\phi}(\rho,\theta)e^{if\Big(\frac{\gamma^0}{2}+\tau_3\Big)}.
\end{equation}
Therefore, the mass term $\Psi^{\dagger}U_{\phi}(\rho,\theta)\Psi$ is invariant under $\phi\rightarrow\phi+f$ and $\Psi\rightarrow\exp if\Big(\frac{\gamma^0}{2}+\tau_3\Big)\Psi$.

In the $\phi=0$ gauge, we can explicitly solve for the low-energy modes bound to the domain wall~\cite{jackiw1976solitons}. 
We start by finding the zero modes of the radial part of the Hamiltonian near the domain wall, and then add the angular kinetic energy and mass as perturbations. 
The radial part of Hamiltonian \eqref{rad ham} is
\begin{equation}
    H_{\rho}=
    \sigma^3\otimes\begin{pmatrix}
        m(\rho)&-v\partial_{\rho}\\
        v\partial_{\rho}&-m(\rho)
    \end{pmatrix}.
\end{equation}
The zero modes satisfy $H_\rho\ket{\phi_\sigma(\rho)}=0$ and are given by
\begin{equation}
    |W_{\sigma}(\rho)\rangle=\frac{G(\rho)}{\sqrt{N}}|\sigma\rangle\otimes 
    \begin{pmatrix}
        1\\
        1
    \end{pmatrix},
\end{equation}
 where $G(\rho)=\exp\Big({-\frac{1}{v}\int_R^\rho}m\Big)$ and $N$ has been chosen such that
\begin{equation}
    \int\mathrm{d}\rho\langle W_{\sigma}(\rho)|W_{\sigma'}(\rho)\rangle=\delta_{\sigma\sigma'}.
\end{equation}
Since, in a general gauge, 
\begin{equation}
    H(\phi)=e^{i\phi(\rho,\theta)\Big(\frac{\gamma^0}{2}+\tau_3\Big)}H(0)e^{-i\phi(\rho,\theta)\Big(\frac{\gamma^0}{2}+\tau_3\Big)},
\end{equation}
the zero modes in a general gauge will be,
\begin{equation}
\label{jackiw-rebbi}
    |W^{\phi}_{\sigma}(\rho,\theta)\rangle=e^{-i\sigma \phi}\frac{G(\rho)}{\sqrt{N}}|\sigma\rangle\otimes
    \begin{pmatrix}
        e^{i\frac{\phi}{2}}\\
        e^{-i\frac{\phi}{2}}
    \end{pmatrix},
\end{equation}
where $|\sigma\rangle=|\pm\rangle$ are the two eigenstates of $\sigma^3$. 
We can obtain the boundary Hamiltonian by projecting the bulk Hamiltonian into the space of the energy modes,
\begin{equation}
    H_{\mathrm{bndy}}=\int q\mathrm{d}\rho\mathrm{d}\theta\langle\psi_{\phi}(\rho,\theta)|H(\phi)|\psi_{\phi}(\rho,\theta)\rangle,
\end{equation}
where $|\psi_{\phi}(\rho,\theta)\rangle=g_+(\theta)|W_+^{\phi}(\rho,\theta)\rangle+g_-(\theta)|W^{\phi}_-(\rho,\theta)\rangle$. 
Note that $\Psi(X)$ is antiperiodic under $2\pi$ shifts of $\theta$, which implies that $g_{\pm}(\theta+2\pi)=-g_{\pm}(\theta)$. 

From the definition of the Jackiw-Rebbi zero energy modes, Eq. \eqref{jackiw-rebbi}, we see that under electromagnetic gauge transformations, $g_{\pm}(\rho,\theta)\rightarrow e^{if}g_{\pm}(\rho,\theta)$. 
However, note that under local rotations the zero modes Eq. \eqref{jackiw-rebbi} transform by shifts of $\phi$, and so the $g_{\pm}(x)$ do not transform under local rotations.
To proceed we introduce the boundary coordinate $\mathrm{d}x=q(R,\theta)\mathrm{d}\theta$ that has the identification $x\sim x+L$, and a two component spinor,
\begin{equation}
    \psi(x)=\begin{pmatrix}
        g_+{(x)}\\
        g_-{(x)}
    \end{pmatrix},
\end{equation}
that is antiperiodic in $x$. we can then write the angular part of the Hamiltonian  as
\begin{equation}
\label{bndyham}
    H_{\mathrm{bndy}}=\int_0^L\mathrm{d}x v\bar{\psi}\tilde\gamma^1\Big(i\frac{\mathrm{d}}{\mathrm{d}x}-\bar{\omega}_x\tilde\gamma^5\Big)\psi.
\end{equation}
In writing Eq. \eqref{bndyham}, we are assuming that $G^2(\rho)$ acts like a projection operator that is strongly peaked at the boundary, in other words,
\begin{equation}
    \int\mathrm{d}\rho Y(\rho,\theta)\frac{2G^2(\rho)}{N}\sim Y(R,\theta),
\end{equation}
for any function $Y(\rho,\theta)$.
We have also introduced a set of boundary gamma matrices, 
\begin{align}
\tilde\gamma^0&=s_y,\;\; \tilde\gamma^1=is_x \nonumber\\ 
\tilde\gamma^5&=\tilde\gamma^0\tilde\gamma^1=s_z,
\end{align} where $s_i$ are Pauli matrices acting on the two component spinor $\psi(x)$. 
Finally, note that
\begin{equation}\bar{\omega}_x=\bar{\omega}_{\theta}\frac{\mathrm{d}\theta}{\mathrm{d}x}=\frac{\bar{\omega}_{\theta}}{q(R,\theta)}=-K_x.\end{equation}

We can also project the bulk mass term into the space of the zero energy modes, and obtain
\begin{flalign}
    H_{\mathrm{mass}}&=\int q\mathrm{d}\rho\mathrm{d}\theta\langle\psi_{\phi}(\rho,\theta)|U_{\phi}(\rho,\theta)|\psi_{\phi}(\rho,\theta)\rangle\nonumber\\&=\int_0^L\mathrm{d}xM\bar{\psi}\psi.
\end{flalign}
Therefore, we see that the boundary Hamiltonian is a $(1+1)$ dimensional massive Dirac fermion coupled to an axial gauge field $\tilde{A}=\bar{\omega}_{x}\mathrm{d}x$. 
We thus see that the mass Eq.~\eqref{massterm} gaps the boundary modes via the coupling to extrinsic geometry, consistent with our expectation of Sec.~\ref{sec:kmatrix}. 
We now further show that the response theory of the boundary is governed by the GJA action.

We quantize the boundary theory by promoting $\psi(x)$ to be an operator with,
\begin{flalign}
    \{\psi_{\sigma}(x),\psi^{\dagger}_{\sigma'}(x')\}=\delta(x-x')\delta_{\sigma\sigma'},
\end{flalign}
and all other anticommutators vanishing.
Since the fermions are antiperiodic in $x$, the delta function is also antiperiodic.
We can eliminate the spin-connection, which is periodic in $x$, by introducing a fermion $\chi(x)$ that is a chiral rotation of $\psi(x)$,
\begin{equation}
\label{chiral_rot}
    \chi(x)=\exp[i\Phi(x)\tilde\gamma^5]\psi(x),
\end{equation}
where $\Phi(x)=\int_0^{x}\bar{\omega}_x$.
The Hamiltonian transforms to,
\begin{flalign}
\label{bndyham2}
    H_{\mathrm{bndy}}&=\int_0^L\mathrm{d}x \bar{\chi}\Big(iv\tilde\gamma^1\frac{\mathrm{d}}{\mathrm{d}x}+M_0(x)+iM_5(x)\tilde\gamma^5\Big)\chi\nonumber\\
    &+\int_0^L\mathrm{d}x\frac{v}{2\pi}\bar{\omega}_x^2,
\end{flalign}
with $M_0(x)=M\cos[2\Phi(x)]$ and $M_5(x)=M\sin[2\Phi(x)]$.
The counterterm in the second line of Eq. \eqref{bndyham2} arises from the need for consistency under iterated chiral transformations \cite{FUJIKAWA_2004}.  
Note that this chiral rotation will change the boundary condition on the fermions as well as the charge density.
The boundary conditions on $\chi(x)$ becomes, $\chi_{\sigma}(x+L)=-e^{i\sigma\Phi(L)}\chi_{\sigma}(x)$, where $\Phi(L)=\int_0^L\bar{\omega}_x$. 
The charge density will change due to the chiral anomaly.
Following Refs. \cite{fujikawa2004path,FUJIKAWA_2004}, if the fermions were coupled to a $U(1)$ gauge field $A$, the change in the path integral measure, under the chiral transformation will be,
\begin{equation}
    \ln J_5(\Phi)=-\frac{i}{\pi}\int\mathrm{d}t\mathrm{d}x\Phi(x)\Big(\partial_0A_x-\partial_xA_0\Big).
\end{equation}
Since $\Phi(x)$ is time independent, the first term is a total derivative, and is zero. 
The second term, after an integration by parts, is proportional to $A_0$,
\begin{equation}
    \ln J_5(\Phi)=-i\int\mathrm{d}t\mathrm{d}xA_0\Big(\frac{\bar{\omega}_x}{\pi}\Big).
\end{equation}
Therefore, the charge density is given by~\cite{FUJIKAWA_2004,stone_1994_bosonization},
\begin{equation}
    j^0(x)=:\chi^{\dagger}\chi:-\frac{1}{\pi}\bar{\omega}_x,
\end{equation}
while the current density remains unchanged. 
Here $:\cdot:$ denotes normal ordering with respect to the ground state of Eq. \eqref{bndyham2} at charge neutrality, i.e. at zero chemical potential and no external fields.

Following Refs. \cite{kane1992transport, von1998bosonization,senechal2004introduction}, we introduce boson fields, $\varphi(x)$ and $\vartheta(x)$ such that,
\begin{equation}
    \begin{pmatrix}
        \chi_1(x)\\
        \chi_2(x)
    \end{pmatrix}=
    \frac{\kappa}{\sqrt{L}}\begin{pmatrix}
        :e^{i[\varphi(x)+\vartheta(x)+\delta_bx]}:\\
        :e^{i[\varphi(x)-\vartheta(x)-\delta_bx]}:
    \end{pmatrix},
\end{equation}
where again, $:\cdot:$ denotes normal ordering with respect to the ground state of Eq.~\eqref{bndyham2} at charge neutrality, $\kappa$ is a Majorana fermion that connects different fermion number sectors\footnote{Comparing with Ref. \cite{rao2023effective}, we see that there are no Klein factors in our bosonization rule since we have chosen to absorb them into the zero mode of $\varphi$.
This does not cause $\vartheta$ to have a nonzero winding since the eigenvalues of the left and right moving momenta are constrained to be integers for the fermions to have antiperiodic boundary conditions, see for example Ref. \cite{stone_1994_bosonization}. 
Thus, they cancel out from the $\vartheta$ mode, which is equal to the sum of the left and right moving bosons.}, and $\delta_b=\frac{\Phi(L)}{L}$.
We have introduced the additional factor of $\delta_bx$ in our bosonization formula following Ref. \cite{von1998bosonization}, so that the bosonized fermions satisfy the twisted boundary conditions after the chiral transformation, $\delta_b$ effectively plays the role of a Fermi energy.
We see that under gauge transformations $\varphi\rightarrow\varphi+f$ while $\vartheta$ is left invariant.
In terms of the boson fields, the kinetic part of the boundary Hamiltonian Eq. \eqref{bndyham2} becomes \cite{kane1992transport},
\begin{equation}
    \int_0^L\mathrm{d}x \bar{\chi}\Big(iv\gamma^1\frac{\mathrm{d}}{\mathrm{d}x}\Big)\chi=\int_0^L\mathrm{d}x\frac{v}{2\pi}\Big((\partial_x\varphi)^2+(\partial_x\vartheta)^2\Big),
\end{equation}
with the fermion charge density being given by,
\begin{equation}
    j^0(x)=\frac{1}{\pi}:\partial_x\vartheta:-\frac{1}{\pi}\bar{\omega}_x.
\end{equation}
Note that the additional factor of $\delta_bx$ in the bosonization formula does not appear in the charge density since it drops out from $\chi^{\dagger}\chi$ after it is normal ordered.  
In order to bosonize the mass terms we will use the following relations for the fermion mass terms,
\begin{equation}
\label{m1}
    \bar{\chi}\chi=\frac{1}{\pi\mu}:\sin2\Big(\vartheta(x)+\delta_bx\Big):,
\end{equation}
and,
\begin{equation}
\label{m2}
    \bar{\chi}\tilde\gamma^5\chi=\frac{i}{\pi\mu}:\cos2\Big(\vartheta(x)+\delta_bx\Big):,
\end{equation}
where $\mu$ is a short distance cutoff scale.
The above relations imply,
\begin{flalign}
\label{mm1}
    M_0(x)\bar{\chi}\chi+iM_5(x)\bar{\chi}\tilde\gamma^5\chi=\frac{M}{\pi\mu}:\sin2\Big(\vartheta(x)-z(x)\Big):,
\end{flalign}
where $z(x)=\Phi(x)-\delta_bx$, is a periodic function, $z(x+L)=z(x)$.
Note that, $\vartheta$ is charge neutral, rotationally invariant and periodic in $x$, and therefore, the mass term, Eq. \eqref{mm1} is well defined.
The full Hamiltonian is,
\begin{flalign}
    H_{\mathrm{bndy}}&=\int_0^L\mathrm{d}x\frac{v}{2\pi}\Big((\partial_x\varphi)^2+(\partial_x\vartheta)^2\Big)\nonumber\\
    &+\frac{M}{\pi\mu}\int_0^L\mathrm{d}x\sin2(\vartheta-z)+\frac{v}{2\pi}\int_0^L\mathrm{d}x\bar{\omega}_x^2.
\end{flalign}
In the saddle point approximation, the mass term pins the value of $\langle\vartheta\rangle$ to $z(x)$ and thus the boundary charge density is given by
\begin{equation}
    \langle\rho(x)\rangle=-\frac{1}{\pi}\bar{\omega}_x+\frac{1}{\pi}\frac{\mathrm{d}z}{\mathrm{d}x}=\frac{2}{2\pi}K_x+\frac{1}{\pi}\frac{\mathrm{d}z}{\mathrm{d}x},
\end{equation}
where we used $\bar{\omega}_x=-K_x$. 
The total charge, $\int\rho(x)$ does not depend on $z(x)$ since $z(x)$ appears in the expression for the charge density as a total derivative and is periodic in $x$.
This boundary charge density implies that the boundary induced action is given by,
\begin{equation}
    S[A]=\frac{\bar{s}}{2\pi}\int_{\partial\mathcal{M}}A\wedge K-\frac{1}{\pi}\int_{\partial\mathcal{M}}z\mathrm{d}A.
\end{equation}
with $\bar{s}=2$. 
The first term is the ``first'' GJA term $S_{\mathrm{GJA}_1}$ of Eq.~\eqref{eq:gjaintro}, while the second term is a nonuniversal local perturbation~\cite{rao2023effective}. 
We note that the second GJA term does not appear in our derivation since $\omega_t=K_t=0$ for the time-independent geometry. 
Thus, consistent with our general argument of Sec.~\ref{sec:kmatrix}, we see that coupling to extrinsic geometry gaps the edge modes for the rotationally invariant higher-order topological insulator, leaving a boundary charge response governed by the GJA action.

\section{Outlook}\label{sec:outlook}

In this work we have systematically examined the edge theory of rotationally invariant topological phases in 2+1 dimensions. 
We showed in Sec.~\ref{sec:kmatrix} on very general grounds that continuous (internal) rotational symmetry provides no extra protection to the edge modes in abelian topological phases. 
As long as charge neutrality (vanishing Hall conductance) and mutual locality (existence of a Lagrangian subgroup) conditions holds for all the fields in the edge theory, we showed how mass terms could be constructed to fully gap the rotationally invariant boundary without breaking any symmetries. 
The mass terms involve vertex operators that explicitly dress the microscopic fields with a coupling to the extrinsic geometry of the boundary. 
Furthermore, we were able to show by explicit integration that, once care was taken to account for local counterterms, the universal terms in the boundary response theory in the presence of the mass terms is governed by the Gromov-Jensen-Abanov action of Eq.~\eqref{eq:gjaintro}. 
This is consistent with the observation of Ref.~\cite{gromov2016boundary} that there is no boundary anomaly associated with continuous rotational symmetry. 
Furthermore, our work generalizes the results of Refs.~\cite{wieder2020strong,manjunath2023rotational,rao2023effective,liu2019shift}, which arrived at a similar conclusion for various models of topological phases protected by discrete rotational symmetry. 
Finally, we applied our approach to analyze the boundary theory for a model higher-order topological insulator first introduced in Ref.~\cite{may2022crystalline}. 
We explicitly solved the Schr\"{o}dinger equation for the boundary modes on a general curved manifold and constructed rotationally invariant mass terms using the boundary tangent and normal vectors. 
We showed via bosonization that the mass terms open a gap on the boundary, and that the charge response of the gapped boundary is governed by the GJA action.

Our work opens up several avenues for further exploration. 
First, our work highlights the importance of extrinsic geometry in understanding the low-energy behavior of topological phases protected by or enriched with spatial symmetries. 
While we focused on continuous rotationally symmetric phases in the present work, our construction of dressed vertex operators and the coupling of microscopic boundary modes to extrinsic geometry can be applied even when continuous rotational symmetry is broken to a discrete subgroup. 
Our formalism points towards a natural way to describe these phases in terms of the spontaneous breaking of continuous rotational symmetry by additional fields in the low-energy effective theory. 
This would allow for the extension of crystalline gauge theory to study boundary modes of topological phases in two dimensions. 
This would also allow for an extension of our work to more exotic nonabelian topological phases.

Second, it would be an interesting future direction to extend this approach beyond two dimensions. 
Recent studies have shown that quasi-two-dimensional topological phases can also host nontrivial response to intrinsic geometric gauge fields, at least when there is a preferred symmetry axis~\cite{may2022topological,may2024crystalline}. 
In these systems too, one should expect that boundary degrees of freedom can couple to extrinsic geometry. 
Additionally, we expect that low-energy degrees of freedom bound to line defects in these phases will be sensitive to the extrinsic geometry of the defect line (such as its curvature and torsion). 
This represents an important avenue for the study of the robust properties of three-dimensional higher-order topological insulators.

Finally, in our work we have focused on the existence of mass terms without regards to their scaling dimensions. 
For the question of the general stability of edge modes this is reasonable: we were primarily concerned with whether an arbitrarily strong external potential could be added to open a gap at the boundary. 
However, we can also ask whether the edge modes are \emph{perturbatively} stable. 
For this we can consider the scaling dimension of the vertex operators Eqs.~\eqref{vertex_op} and \eqref{vertex_op_2} that build up the mass terms. 
Power-counting renormalization of the boundary action shows that if the scaling dimension $\Delta_i$ for each vertex operator is smaller than $2$, the operator will be relevant and so any infinitesimal mass perturbation will grow under renormalization, opening a gap at the boundary. 
In general, the scaling dimension of the vertex operators depends on $\Lambda_i, K,$ and the velocity matrix~\cite{kane1995impurity}. 
We will leave a detailed study of the perturbative stability of the edge modes to future work. 
However, it is instructive to examine the renormalization group flow for the mass Eq.~\eqref{bulkmass} in our concrete example from Sec.~\ref{sec:example}. 
We see from Eq.~\eqref{mm1} that the relevant vertex operator generating the mass term is $V\sim :e^{2i\vartheta}:$, which has scaling dimension $\Delta=2$. 
This means the mass term in this rotationally invariant higher-order topological insulator is marginal: whether or not a given realization of this phase has a gapped or gapless boundary will depend on nonuniversal details beyond the $\mathbf{k}\cdot\mathbf{p}$ approximation in Eq.~\eqref{k.p ham}. 
This highlights the fact that the boundary properties of higher-order topological insulators may depend on nontopological details~\cite{lin2022spin}, which is an important avenue for further research.

\begin{acknowledgments}
The authors thank T.~L.~Hughes and J.~May-Mann for helpful discussions. 
The authors would also like to thank M. 
Stone for discussions and for sharing his extremely comprehensive pedagogical lecture notes. 
This work was supported by the National Science Foundation under grant No.~DMR-1945058.
\end{acknowledgments}
\appendix
\section{Chern-Simons boundary theory with external gauge fields}
\label{app:A}
In this appendix we show how to obtain the boundary theory for a general $K$-matrix Chern-Simons theory coupled to electromagnetic and rotational gauge fields. 
We start with the bulk Chern Simons action given in condensed form by
\begin{equation}
\label{ACS}
    S=\int \mathrm{d}^3X\Big[-\frac{K^{IJ}}{4\pi}\epsilon^{\mu\nu\sigma}a^I_{\mu}\partial_{\nu}a^J_{\sigma}-\frac{t^I_a}{2\pi}\epsilon^{\mu\nu\sigma}a^I_{\mu}\partial_{\nu}A^a_{\sigma}\Big].
\end{equation}
The index $a$ runs over the EM field $A_{\mu}$ and the spin connection $\omega_{\mu}$. 
The coupling to the current, $J^{\mu}_I=\frac{1}{2\pi}\epsilon^{\mu\nu\sigma}\partial_{\nu}a_{\sigma}^I$ is given by $t^I_aJ^{\mu}_IA^a_{\mu}$, but we integrated it by parts to make the action manifestly gauge invariant in $A^a$.
In other words, we have added a boundary term, 
\begin{equation}\label{eq:gaugeboundaryterm}
    \int_{\mathcal{M}}\frac{t^I_a}{2\pi}\mathrm{d}\Big(A^a\wedge a^I\Big)=\int_{\partial\mathcal{M}}\frac{t^I_a}{2\pi}\Big(A^a\wedge a^I\Big)\Big|_{\partial\mathcal{M}}
\end{equation}
to the action Eq. \eqref{s(a,w)action} to get Eq. \eqref{ACS}.
This makes explicit that the degrees of freedom of the theory are gauge transformations of $a^I$ which do not vanish at the boundary. 
The action \eqref{ACS} is metric-independent as a classical theory on a manifold with no boundary, but in the full quantum mechanical theory, the metric enters the path integral through the gauge fixing term \cite{witten1989quantum,gromov2014framinganomaly,Bar-Natan:1991fix}, and, on a manifold with a boundary, through the boundary conditions. 
The gauge fixing term leads to the gravitational anomaly term in the effective theory, which does not appear due to our assumption that the chiral central charge is zero.
The boundary conditions, which we choose to be,
\begin{equation}
\label{generalV_bc}
    \left. a_0^I\right|_{\partial\mathcal{M}}=\frac{1}{\sqrt{\gamma}}\left. 
    V^I_{\hphantom{I}J}a_x^J\right|_{\partial\mathcal{M}},
\end{equation}
for a symmetric matrix $V$ and boundary spatial metric $\gamma$, introduce metric dependence into our boundary theory.
This boundary condition leads to a well defined variational principle if the symplectic 2-form,
\begin{equation}
    \omega=\frac{1}{4\pi}\int_{\partial\mathcal{M}}K^{IJ}\delta a_I\wedge\delta a_J
\end{equation}
vanishes when restricted to fields that obey the boundary conditions \cite{fliss2017interface}.
This holds for our choice of boundary conditions if $V$ commutes with $K$. 
Our choice of boundary conditions Eq.~\eqref{generalV_bc} ensures that the resulting theory remains invariant under spatial diffeomorphism $x\rightarrow x'(x)$ on the boundary and reduces the Chern-Simons path integral to that of a theory of bosons with spacetime-dependent velocities. 
Note that one can consider more general topological boundary conditions~ \cite{kapustin2011topological} that do not depend on the choice of metric, but then the nonuniversal dynamics of the boundary theory need to be treated with more care.

We will split Eq. \eqref{ACS} into space and time components,
\begin{widetext}
\begin{flalign}
\label{eq:bulkactionstart}
    S=\int_{\mathcal{M}}\mathrm{d}^3X\epsilon^{ij}\Big[&-\frac{K^{IJ}}{4\pi}\Big(a_0^I\partial_ia_j^J-a_i^I\partial_0a_j^J+a_i^I\partial_ja_0^J\Big)
    -\frac{t^I_a}{2\pi}\Big(a_0^I\partial_iA^a_j-a_i^I\partial_0A_j^a+a_i^I\partial_jA_0^a\Big)\Big],
\end{flalign}
\end{widetext}
and integrate the third term by parts, using the boundary condition Eq. \eqref{generalV_bc} .
Since $a_0^I$ does not appear with time derivatives, it imposes a constraint on the gauge fields,
\begin{equation}
    K^{IJ}f_{ij}^J=-t^I_aF^a_{ij}\implies a_i^I=\partial_i\phi^I-(K^{-1})^{IJ}t_J^aA_i^a.
\end{equation}
Plugging this back in, and integrating by parts, we get that the bulk action becomes,
\begin{equation}
\label{buulk}
    S_{\mathrm{bulk}}=\frac{\sigma_{ab}}{4\pi}\int_{\mathcal{M}}A^a\wedge\mathrm{d}A^b,
\end{equation}
where $\sigma_{ab}=t^T_aK^{-1}t_b$.
Collecting the boundary terms that arise from integrating Eq.~\eqref{eq:bulkactionstart} by parts and combining them with the boundary term Eq.~\eqref{eq:gaugeboundaryterm}, we arrive at the boundary boson theory
\begin{widetext}
\begin{flalign}
    S_{\mathrm{bndy}}=-\int_{\partial\mathcal{M}}\mathrm{d}t\mathrm{d}x\Big[&\frac{K^{IJ}}{4\pi}\partial_x\phi^I\partial_t\phi^J-\frac{(KV)^{IJ}}{4\pi\sqrt{\gamma}}\partial_x\phi^I\partial_x\phi^J-\frac{\partial_x\phi^I}{2\pi}\Big(t^a_IA^a_t-\frac{1}{\sqrt{\gamma}}(Vt_a)_IA_x^a\Big)\Big. \nonumber \\
    &+\frac{\sigma_{ab}}{4\pi}A_t^aA_x^b-\frac{\sigma'_{ab}}{4\pi\sqrt{\gamma}}A_x^aA_x^b\Big],
\end{flalign}
\end{widetext}
where $\sigma'_{ab}=t^T_a(K^{-1}V)t_b$, and the additional minus sign is due to the orientation of the normal vector to the boundary (see Ref. \cite{rao2023effective} for details). 
Once we plug in $t_{A}^I=t^I$ and $t_{\omega}^I=s^I$, and integrate the $\omega\wedge\mathrm{d}A$ term in \eqref{buulk} by parts, we get Eq. \eqref{bndy}. 
Note that, for the boundary theory to have a positive Hamiltonian, $KV$ must be a positive definite matrix.
\begin{widetext}
\section{Proof of gapped boundary action}
\label{app:B}
In this appendix, we show how to explicitly integrate out the boundary bosons in Eq.~\eqref{bndy} in the presence of the mass terms Eq.~\eqref{massterm} to obtain the GJA action. 
Plugging the field decomposition Eq. \eqref{decom} into the boundary action action Eq. \eqref{bndy} and adding the mass terms, we get

\begin{flalign}
\label{eta,betaeq}
    S_{\mathrm{bndy}}=-\int_{\partial\mathcal{M}}\mathrm{d}t\mathrm{d}x\Big[&\frac{B^{IJ}}{4\pi}\partial_x\beta_I\partial_t\beta_J+\frac{A^{IJ}}{2\pi}\partial_t\beta_J\partial_x\eta_I-\frac{v}{4\pi}\Big((\partial_x\beta^I)^2+(\partial_x\eta^I)^2\Big)\Big. \nonumber \\
    &-\frac{1}{2\pi}A_tt^I\partial_x\beta^I-\frac{\partial_x\beta^I}{2\pi}\Big(s_2^I\omega_t-vl_2^I\omega_x\Big)+\frac{v}{2\pi}A_xq^I\partial_x\eta^I-\frac{\partial_x\eta^I}{2\pi}\Big(s_1^I\omega_t-vl_1^I\omega_x\Big)\Big. \nonumber \\
    &-v\frac{\sigma'}{4\pi}A_x^2+\frac{\bar{s}}{2\pi}\omega_xA_t-v\frac{s'}{2\pi}\omega_xA_x-v\frac{l_s'}{4\pi}\omega_x^2+\frac{l_s}{4\pi}\omega_t\omega_x+\mathcal{L}_\mathrm{mass}\Big]
\end{flalign}
where we have used that $q^T\beta(t,x)=t^T\eta(t,x)=0$ and $K$ is of the form \eqref{Kmat}. 
We see from the action that $\eta_I$ and $\frac{A^{IJ}}{2\pi}\beta_J$ are canonically conjugate variables. 
Since the action for $\eta_I$ is quadratic, we can integrate it out by substituting the equations of motion.
The equation of motion of $\eta_I$ is,
\begin{align}
\label{eom}    \partial_x(v\partial_x\eta_I)&=A_{IJ}\partial_x\partial_t\beta^J+q_I\partial_x(vA_x)\nonumber \\
&-(s_1)_I\partial_x\omega_t+(l_1)_I\partial_x(v\omega_x).
\end{align}
We solve for $\partial_x\eta_I$ by removing a derivative from Eq. \eqref{eom} and then we plug it back into Eq. \eqref{eta,betaeq}.
Finally, using 
\begin{align}
\bar{s}&=t^TK^{-1}s=t^TK^{-1}s_1, \\ s'&=t^TK^{-1}l=t^TK^{-1}l_1,
\end{align}
we get Eq. \eqref{gapped}.
Note that we have assumed $v\neq 0$ in Eq. \eqref{eom}.
Otherwise, $\partial_x\eta^I(x)$ sets the following constraint,
\begin{equation}
    A^{IJ}\partial_t\beta_J-(s_1)^I\omega_t=0.
\end{equation}
This makes the third and fourth terms zero in Eq. \eqref{GJA}.
Next, using the mass term to pin the value of $\langle\beta^I(x)\rangle$ to $l_2^I\alpha(x)$, we see that the boundary response is
\begin{equation}
    S_{\mathrm{bndy}}=\int_{\partial\mathcal{M}}\mathrm{d}t\mathrm{d}x\Big[\frac{\bar{s}}{2\pi}\Big(K_xA_t-K_tA_x\Big)-\frac{s_1^Ts_1}{4\pi v}K_t^2+v\frac{l_2^Tl_2}{4\pi }K_x^2+\frac{1}{4\pi}(l^T_1s_1-l^T_2s_2)K_xK_t+\frac{l_s}{4\pi}\Big(K_x\omega_t-K_t\omega_x\Big)\Big],
\end{equation}
where we used 
\begin{equation}
l_2^TBl_2=l^T_2s_2-l_2^TA^Tl_1=l^T_2s_2-l_1^TAl_2=l^T_2s_2-l_1^Ts_1.
\end{equation}

Thus, we obtain the GJA term corresponding to the second Wen-Zee term as well.
\end{widetext}
\bibliography{refs.bib}

\end{document}